\documentclass[pdflatex,sn-mathphys-num]{sn-jnl}

\usepackage{graphicx}%
\usepackage{multirow}%
\usepackage{amsmath,amssymb,amsfonts}%
\usepackage{amsthm}%
\usepackage{mathrsfs}%
\usepackage[title]{appendix}%
\usepackage{xcolor}%
\usepackage{textcomp}%
\usepackage{manyfoot}%
\usepackage{booktabs}%
\usepackage{algorithm}%
\usepackage{algorithmicx}%
\usepackage{algpseudocode}%
\usepackage{listings}%

\begin{document}

\title[Effectiveness of Some 0 dB Cryogenic Microwave Attenuators as Thermal Heatsinks]{Effectiveness of Some 0 dB Cryogenic Microwave Attenuators as Thermal Heatsinks}

\author*[1]{\fnm{J.T.} \sur{Klomp}}\email{jklomp@purdue.edu}

\author[1]{\fnm{Haoyun} \sur{Huang}}

\author[1]{\fnm{G.A.} \sur{Cs\'athy}}

\affil*[1]{\orgdiv{Department of Physics and Astronomy}, \orgname{Purdue University}, \orgaddress{West Lafayette, IN 47907, USA}}

\abstract{Coaxial cables are widely used in radiofrequency and microwave cryogenic setups for condensed matter and quantum experiments. Since the inner conductor of coaxes is often in good thermal contact with the sample to be measured, it is desirable to know the phononic heat channeled by the inner conductor. Although cryogenic attenuators are widely used to thermalize the inner conductor of coaxes, to our knowledge, quantitative information is not available. We present data on the effectiveness of three commercially available 0 dB attenuators as thermal heatsinks. In particular, we measured the temperature of the inner pin of several 0 dB attenuators under a heat load. This information will aid in designing and carefully controlling the thermal environment of samples in high-frequency experiments mounted in dilution refrigerators and on nuclear demagnetization stages.}


\maketitle

\section{Introduction}\label{sec1}

Experiments on qubits \cite{bouchiat_quantum_1998, friedman_quantum_2000, martinis_rabi_2002, koch_charge-insensitive_2007, elzerman_single-shot_2004} as well as on various condensed matter systems \cite{paalanen_electrical_1992, ye_correlation_2002, friess_negative_2017} call for measurements in the radiofrequency and microwave regimes at millikelvin temperatures. These measurements use cryogenic coaxial cables to transmit signals in and out of the refrigerator.
It is widely appreciated that coaxial cables heat the various stages of a cryogenic instrument. For example, heating from coaxes causes an increase of the temperature of the mixing chamber of a dilution refrigerator and that of the sample to be measured. 
Controlling such heating effects is important for low temperature measurements, especially those that focus on the ultra-low temperature regime \cite{pan_exact_1999, xia_ultra-low-temperature_2000, clark_method_2010, samkharadze_integrated_2011, iftikhar_primary_2016, palma_-and-off_2017, yurttagul_indium_2019, wang_piezo-driven_2019, jones_progress_2020, levitin_cooling_2022}. 

Numerous properties of cryogenic high frequency circuits are well-understood.
For example, there is ample information available on the thermal conduction of various cryogenic coaxes \cite{kushino_thermal_2005,kushino_thermal_2018,krinner_engineering_2019}.
Furthermore,  mitigating the black body radiation that impinges on the sample with cryogenic microwave attenuators \cite{santavicca_impedance-matched_2008, thalmann_comparison_2017, yeh_microwave_2017, yeh_hot_2019, danilin_engineering_2022}
or with infrared filters \cite{martinis_experimental_1987, vion_miniature_1995, zorin_thermocoax_1995, fukushima_attenuation_1997, bladh_comparison_2003, milliken_50_2007, lukashenko_improved_2008, scheller_silver-epoxy_2014} is well documented.

In contrast, quantitative knowledge on the thermalization of the inner conductor of coaxes is lacking. Indeed, it is widely appreciated that
the inner conductor of coaxes is in 
poor thermal contact with the outer conductor.
For the most common coaxial design, poor thermalization of the inner conductor is due to poor thermal conductivity of the teflon spacers. In addition, differential thermal contraction effects may result in the formation of
a gap between the teflon and metallic parts of the coax. 
The sample to be measured is typically in good thermal contact with the inner conductor of the coax. Therefore, the sample temperature will be adversely affected by the poor thermalization of the inner conductor.

Microwave attenuators with various attenuation levels are
commonly used as thermal anchors of the inner conductor of coaxes. However, to our knowledge, the effectiveness of such thermal anchors is not known. In this paper, we provide quantitative information on the thermal anchoring properties of three commercially available 0~dB attenuators.

\section{Methods}\label{sec2}

\begin{figure}[b]
\centering
\includegraphics[width = 2.5 in]{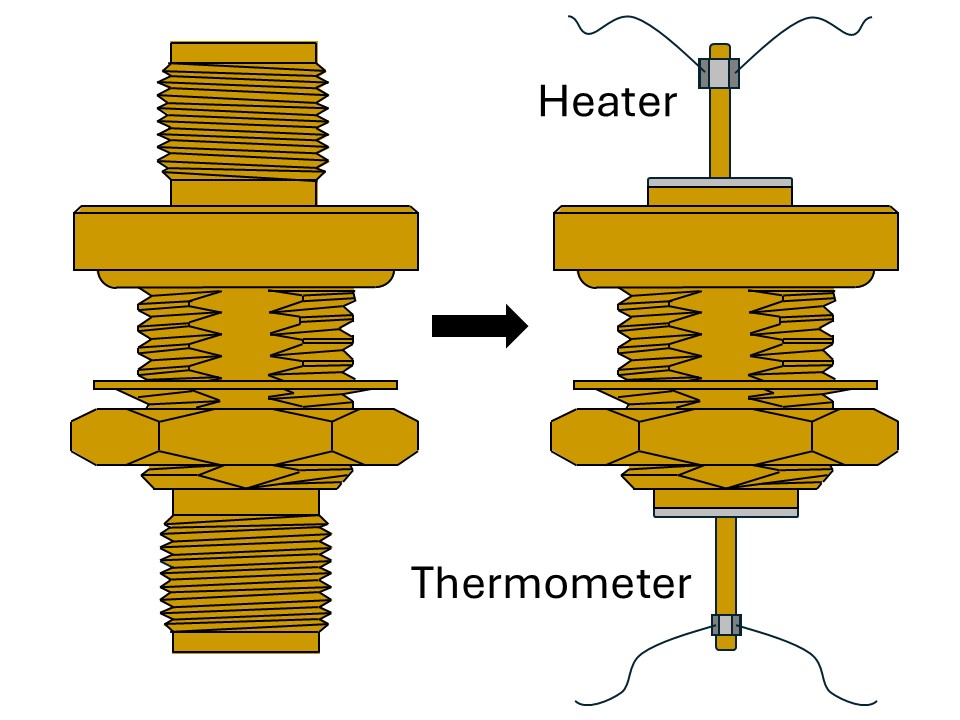}
\caption{\label{Fig1} In order to expose the inner pin of a female SMA connector, its threads were filed down and the teflon spacer was removed. The diagram illustrates this for the 0~dB attenuator ATT-1.
The chip heater and thermometer attached to the inner pin are also shown.}
\end{figure}

In this study, three commercially available 0~dB attenuators were compared. Each attenuator comes with SMA connectors.
ATT-1 is from Fairview Microwave \cite{ATT1}; it has a hermetically sealed design. Such attenuators use specialized sealing techniques, such as glass-to-metal or epoxy seals, to join the inner pin to the body of the attenuator, ensuring thereby a direct thermal contact between the inner and the outer conductor. ATT-2, from Amphenol XMA Corp. \cite{ATT2}, has a teflon spacer between the inner and outer conductor. ATT-3 \cite{ATT3}, from Quantum Microwave, uses a non-superconducting microstrip line deposited on a quartz crystal. For ATT-3, the quartz crystal provides enhanced thermal contact with the body of the attenuator. Not all of these attenuators are intended for cryogenic applications. ATT-1 is specified by the manufacturer to work down to an operating temperature of -65$^\circ$C, whereas ATT-2 and ATT-3 have minimum operating temperatures of 4 K and 10 mK respectively.

The inner pin of a female SMA connector is typically recessed in a teflon insulator, which prevents the fitting of a heater or thermometer necessary for thermal characterization. To gain access to the inner pin, we first remove the outer threaded part of the connector by filing it off with a miniature file. We then remove the teflon insulation. At this point, the inner pin is fully exposed, as shown for ATT-1 in Fig.\ref{Fig1}. We proceed by attaching a heater chip \cite{Heater} with GE varnish to one exposed pin and a chip thermometer \cite{Thermometer} to the other pin. The heater is a metal thin film of size 0805. The thermometer is a ruthenium oxide chip of size 0603 \cite{ myers_ruthenium_2021}; this thermometer was calibrated against the mixing chamber thermometer. These chips are not in electrical contact with the inner pin of the connector. Here we are using the Electronic Industries Alliance (EIA) standards for chip sizes.
Manganin wires of 25 $\mu$m diameter and $\sim$4 cm length were used to connect the heater and thermometer to the measurement lines. It was determined that the thermal conductivity of the manganin wires had an impact on our results on the order of 1\%.


As shown in Fig.\ref{Fig1}, we apply a heating power $P$ to one end of the exposed pin of the connector and measure the temperature $T$ of the thermometer attached to the other end of the pin. We opted for this measurement setup because of the geometrical layout of the attenuator as a heatsink: the heat current flows into the top pin of the attenuator from a warmer stage of the refrigerator, and we are interested in the temperature of the bottom pin, i.e. the other end of the attenuator. The body of the attenuator is in strong thermal contact with the mixing chamber of the refrigerator, of temperature $T_{mx}$.

The thermal response is the ratio of $T - T_{mx}$ and $P$ 
\begin{equation}\label{thermal transresistance}
    R(T_{mx}) = \frac{T - T_{mx}}{P}.
\end{equation}
Here, $T - T_{mx}$ is small compared to $T_{mx}$, typically less than $20\%$. For larger $T - T_{mx}$, i.e. for larger values of $P$, one obtains a $P(T)$ curve. The two quantities are related by
\begin{equation}\label{heatload}
    P\left(T\right)=\int^T_{T_{mx}} \frac{1}{R(\mathcal{T})}\,d\mathcal{T} 
\end{equation}

By adopting the language of two-port networks, $R$ is the thermal transresistance. In this language, to obtain the thermal resistance, the heater and thermometer need to be attached to the same pin of the attenuator (not the configuration of our setup, so not shown in Fig.\ref{Fig1}). Although the thermal resistance measured this way is useful, we opt for measuring the thermal transresistance
because it is more relevant to the effectiveness of thermal anchoring. 
We note that for the simple thermal model in which the thermal conductivity between the upper pin and the lower pin is infinite, 
the thermal transresistance is identical to the thermal resistance. In contrast,
for a thermal model with a distributed thermal resistance between the upper and lower pins, applicable for example for ATT-3, the two quantities will be different.

~\\

\section{Results \& Conclusion}

\begin{figure}[t]
\centering
\includegraphics[width = 3.5 in]{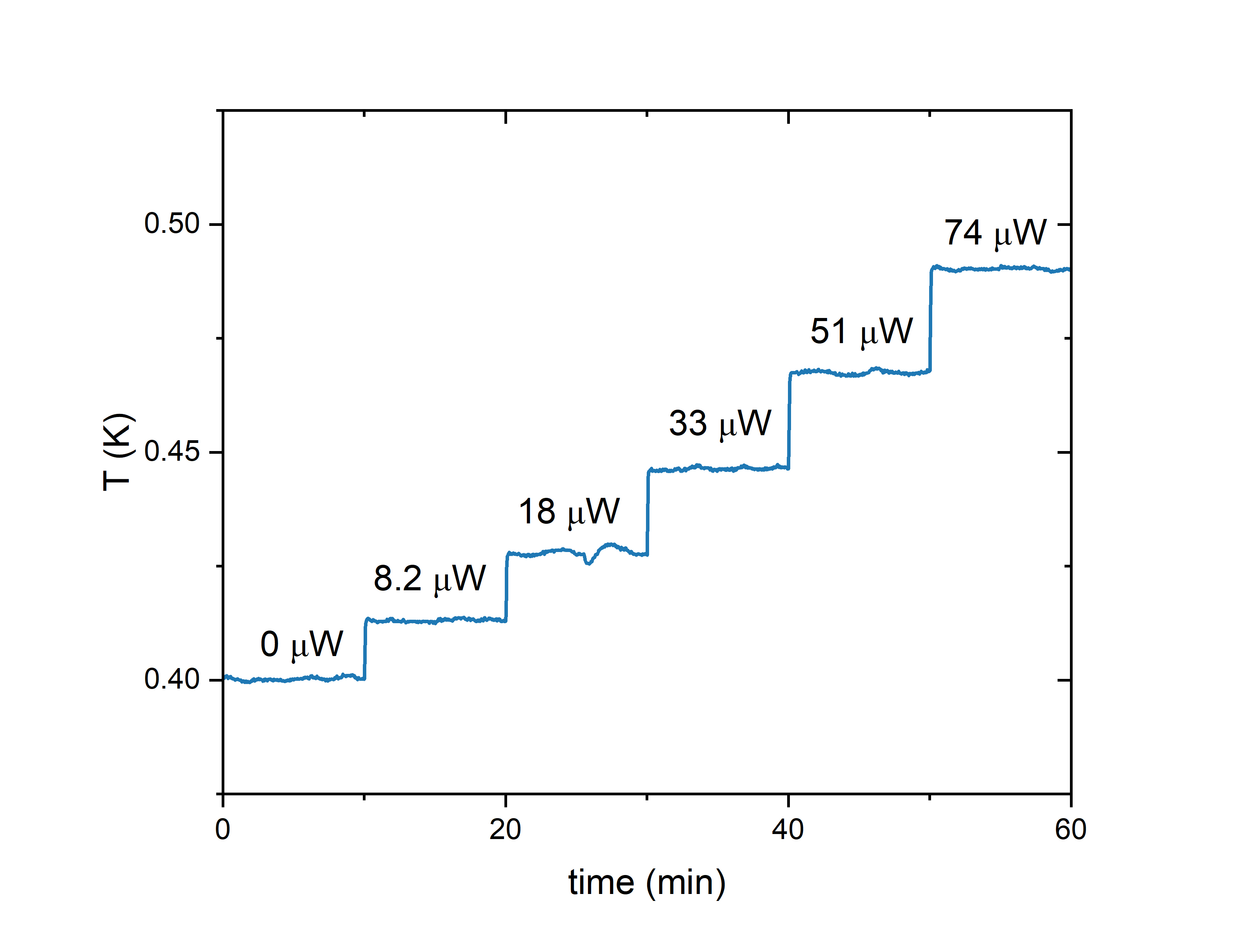}
\caption{\label{Fig2} Readings of the thermometer attached to ATT-3 as the heating power $P$ is increased. Numbers indicate the value of the applied power $P$. The body of the attenuator is held at $T_{mx} = 400$~mK.}
\end{figure}

\begin{figure}[b]
\centering
\includegraphics[width = 1\textwidth]{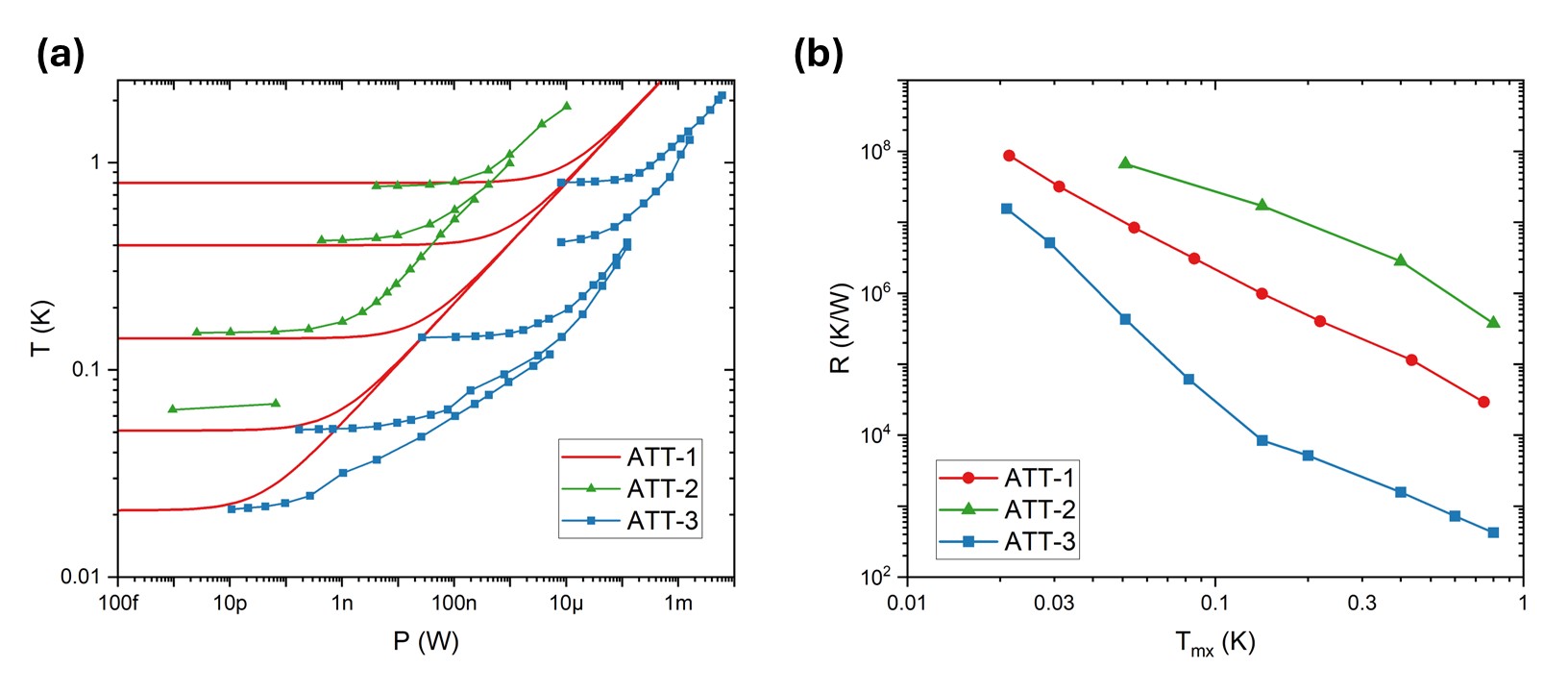}
\caption{\label{Fig3} (a) Thermal response curves of the three attenuators. (b) The thermal transresistance $R$ plotted against mixing chamber temperature $T_{mx}$ for the three attenuators.}
\end{figure}

During the measurements, the mixing chamber temperature $T_{mx}$ was controlled with a PID feedback loop such that it stayed constant. Power was applied incrementally to the chip heater, and the resulting temperature of the chip resistor was recorded. A typical data set is shown in 
Fig.\ref{Fig2} for the ATT-3 attenuator.
For ATT-1, we measured $R(T_{mx})$ at several temperatures according to Eq. \ref{thermal transresistance}, fit the data shown in Fig. \ref{Fig3}b to a power law, and used Eq. \ref{heatload} and the extracted fitting function to obtain $P(T)$ curves. For ATT-2, and ATT-3 we measured $P(T)$ curves at several $T_{mx}$ and obtained $R(T_{mx})$ from Eq. \ref{thermal transresistance}. In both cases to obtain $R(T_{mx})$, $T - T_{mx}$ was at most $20\%$ of $T_{mx}$.

The $P(T)$ curves and thermal transresistance $R$ of the three attenuators are presented in Fig.\ref{Fig3}a and Fig.\ref{Fig3}b , respectively. 
These results show that the thermal performance of various microwave attenuators varies significantly and demonstrate the importance of characterizing them for use in cryogenic applications.
Notably, ATT-3 is the most efficient thermal anchor among the three attenuators we measured as it can tolerate larger power levels, and it has the lowest thermal transresistance at a given temperature. It is worth pointing out that ATT-3 had the lowest specified operating temperature, and it was the only attenuator measured in its operating temperature range guaranteed by the manufacturer.

\begin{figure}[b]
\centering
\includegraphics[width = 3.5 in]{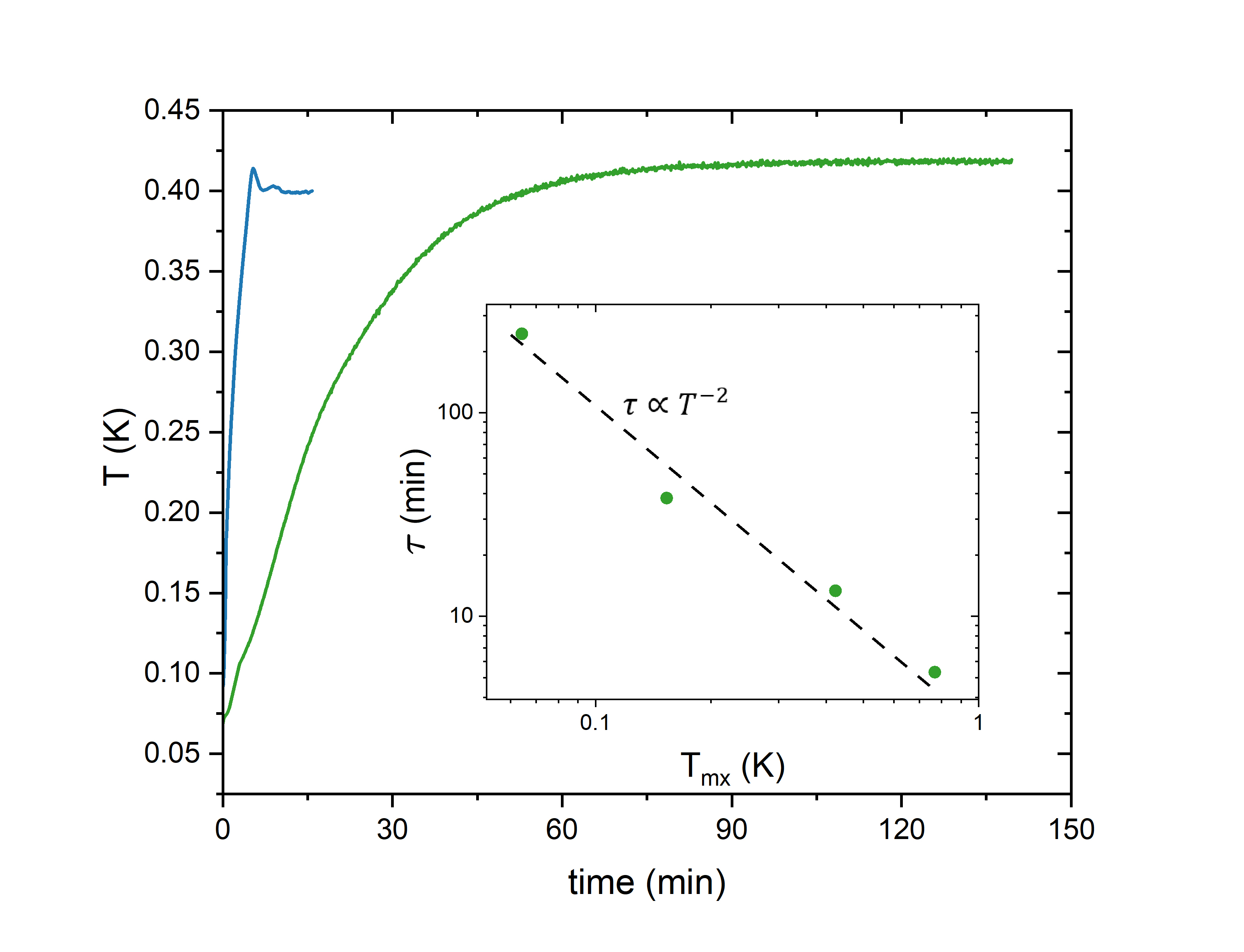}
\caption{\label{Fig4} Temporal responses of ATT-2 (green trace) and ATT-3 (blue trace) from changing the mixing chamber from $T_{mx}=75$~mK to $T_{mx} = 400$~mK. The slight oscillation at the peak of the blue curve is due to overshoot in the PID feedback loop. Inset shows the temperature dependence of the thermal time constant $\tau$ for ATT-2. The dashed line is a guide to the eye.}
\end{figure}

The weakest thermal anchor is ATT-2. A weak thermal contact will result in long thermal equilibration times.  
The heat capacity of the attenuator and of the added thermometer and heater chips plays a significant role in the equilibration time of the attenuator. Assuming that $C\propto  T$ and $R\propto T^{-3}$, the time constant $\tau=RC$ follows an inverse square law in temperature which is consistent with our data.
Fig.\ref{Fig4} illustrates the difference in the equilibration times for ATT-2 and ATT-3
measured at 400~mK. In both cases, the PID parameters are set such that the mixing chamber stabilized after ten minutes.
The inset of Fig.\ref{Fig4} shows the strong temperature dependence of the thermal time constant $\tau$ as a function of temperature. The thermal time constant $\tau$ was obtained from exponential fits to the time dependence of the temperature.

In conclusion, we characterized the thermal response of three commercially available 0~dB attenuators. Such devices are useful for thermal anchoring of coaxial cables and thermalizing the inner conductor of the coax. We found that there is a significant difference in the performance of the three attenuators studied; the most effective is based on a quartz crystal. We also reported long thermal equilibration times for the inner pin of the teflon-based attenuator. The measured thermal response curves will aid in assessing the thermal environment of samples in high-frequency experiments and will be useful especially for experiments performed in the ultra-low temperature regime.

\section*{Acknowledgment}
\subsection*{Funding}

This work was supported by the US Department of Energy Basic Energy Sciences Program under the award DE-SC0006671.

\bibliography{sn-bibliography}

\end{document}